\documentclass[aps,pra,showpacs,twocolumn]{revtex4}

\usepackage[dvips]{graphicx}
\usepackage{amsmath}

\begin{document}

\title{Coherent excitation  of a single atom to a Rydberg state}

\author{Y. Miroshnychenko,$^1$ A. Ga\"{e}tan,$^1$ C. Evellin,$^1$ P. Grangier,$^1$ D. Comparat,$^2$ P. Pillet,$^2$ T. Wilk,$^1$ and A. Browaeys$^1$}
\affiliation{$^1$Laboratoire Charles Fabry, Institut d'Optique, CNRS, Univ Paris-Sud, Campus Polytechnique, RD 128, 91127 Palaiseau cedex, France\\ $^2$Laboratoire Aim\'{e} Cotton, CNRS, Univ Paris-Sud, B\^{a}timent 505, Campus d'Orsay, 91405 Orsay
cedex, France}

\date{\today}

\begin{abstract}
We present the coherent excitation of a single Rubidium atom to the Rydberg state $58d_{3/2}$ using a two-photon transition. The experimental setup is described in detail, as well as experimental techniques and procedures. The coherence of the excitation is revealed by observing Rabi oscillations between ground and Rydberg states of the atom. We analyze the observed oscillations in detail and compare them to numerical simulations which include imperfections of our experimental system. Strategies for future improvements on the coherent manipulation of a single atom in our settings are given.
\end{abstract}



\maketitle

\section{Introduction}

Rydberg atoms have attracted a lot of research interests in the last decades
thanks to their exaggerated properties~\cite{Gallagher94}. Among these properties is their large interaction strength, which is enhanced by many orders of magnitude compared to ground state atoms. This long-range interaction  between two Rydberg atoms has led to a wealth of studies. For example it was suggested~\cite{Greene00, Boisseau02} and recently  demonstrated~\cite{Farooqi03, Bendkowsky09, Overstreet09} that two or more Rydberg atoms or Rydberg and ground state atoms can be bound into exotic giant molecules using photo-association techniques. Another example where the strong interaction plays a role is the Rydberg blockade, an effect where, in a volume smaller than the interaction range, only one atom of an ensemble can be excited into a Rydberg state.

Until recently, the excitation of atoms to Rydberg states was performed using non-coherent laser excitation, usually in multiple steps. This technique led, for example, to the observation of the Rydberg blockade in clouds of ultra-cold atoms~\cite{Tong04, Singer04, Afrousheh04, Cubel05, Vogt06}. In contrast, the coherent excitation of atoms to Rydberg states has been demonstrated only in recent years, starting with experiments using cold thermal clouds of atoms~\cite{Deiglmayr06, Reetz08} and Bose-Einstein  condensates~\cite{Heidemann07, Heidemann08, Raitzsch08}. It was followed by the demonstration of the coherent Rydberg excitation of individual atoms trapped in tight optical dipole traps~\cite{Johnson08, Zuo09}. Along these lines, electromagnetically induced transparency has been reported in room temperature atomic vapor~\cite{Mohapatra07} and microcells~\cite{Kuebler10}, as well as in cold atomic clouds~\cite{Weatherill08}, where also coherent population trapping~\cite{Schempp10} has been observed recently. All those demonstrations were also made possible by the technical development of easy to handle laser light sources, especially in the blue region of the electromagnetic spectrum,  that provide a sufficient amount of power and good stability.

This progress in the coherent manipulation of atoms between ground and Rydberg states is accompanied by an increased interest on the theoretical side. It was  recognized that Rydberg atoms can play an important role in the field of quantum information processing, using either atomic ensembles or arrays of individual atoms~\cite{Jaksch00, Lukin01, Moller08, Mueller09} (see  also~\cite{Saffman10} for a recent review). In these proposals, stable ground  states of single atoms are used as qubit states \cite{Kuhr03, Yavuz06, Jones07} and the controlled coherent interaction between the qubits can be realized via an auxiliary Rydberg state. More generally the long-range interaction between atoms can be tailored and  used to engineer quantum states of an ensemble of atoms or to perform quantum simulations~\cite{Weimer10}. A main ingredient of those proposals is the coherent excitation of an atom to a Rydberg state and the Rydberg blockade.

Recently the coherent excitation of individual atoms trapped in tight optical dipole traps led to the observation of controlled interactions between two individual Rydberg atoms by two groups~\cite{Urban09,Gaetan09}. This work was followed by the demonstration of the entanglement of two atoms~\cite{Wilk10} and of a Controlled NOT gate~\cite{Isenhower10}. The aim of the present paper
is to present in detail the coherent excitation of individual atoms that we used in experiments reported in Ref.~\cite{Gaetan09,Wilk10}. In particular we describe the details of the experimental setup and techniques, and investigate the performance of the experiment.

The paper is organized as follows. In Section~\ref{section_setup}, we present the experimental setup and the laser system used to excite the atoms in the Rydberg state. We also detail the experimental procedure used to align the lasers and to detect the Rydberg excitation of the atom.  Section~\ref{section_spectroscopy}  deals with the Rydberg state spectroscopy on a single atom with some emphasis on the level $58d_{3/2}$. In  Section~\ref{section_Rabi}, we present Rabi oscillations between the ground and the Rydberg state of a single atom. Moreover, we compare these measurements with numerical calculations from a model which includes our independently measured experimental imperfections.

\section{Experimental setup and techniques}\label{section_setup}

In this section we present briefly the experimental setup and the trapping of single $^{87}$Rb atoms in optical tweezers which has been described in earlier publications \cite{Schlosser01, Gaetan09, Wilk10}. We then detail  the laser system used for the excitation of the atom from the ground state to a Rydberg state. We present the alignment technique of the laser beams on the atom. As an aside,  we use this procedure  to estimate the photo-ionization cross-section from the state $5p_{3/2}$. Finally, we discuss the detection of a successful excitation of the atom in the Rydberg state.

\subsection{Single atom trapping}
A single $^{87}$Rb atom is held in an optical tweezers which is formed by a 810~nm laser beam focused with a high numerical aperture lens to a waist of $0.9~\mu$m \footnote{All beam waists in this letter are given at a $1/e^2$-waist radius.}. We operate the trap at a typical power of $0.5$~mW which corresponds to a trap depth of $0.6~$mK. The trap is loaded from an optical molasses created by three retro-reflected cooling beams with a power of about 3~mW per beam collimated to a waist of $1.8$~mm. The detuning of the cooling laser from the $(5s_{1/2}, F=2)$ to $(5p_{3/2}, F=3)$ transition is $-5\Gamma$, where $\Gamma/2\pi =5.75$~MHz is the decay rate of the $5p_{3/2}$ level. A repumping beam drives the $(5s_{1/2}, F=1)$ to $(5p_{3/2}, F=2)$ transition.

\begin{figure}
\includegraphics{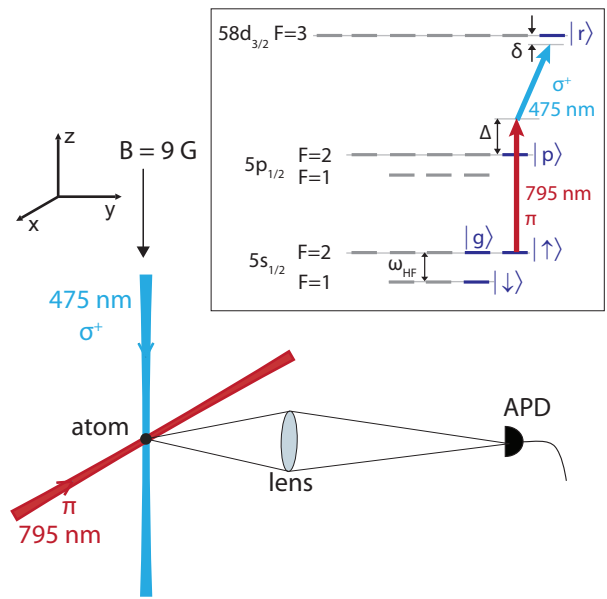}
\caption{\label{fig:setup} 
(Color online) Scheme of the experimental setup. A single atom is trapped in a dipole trap and cooled by optical molasses (not shown). The atomic fluorescence light is detected on a single-photon counting avalanche photo-diode (APD) allowing to detect the presence or absence of an atom in the dipole trap. A blue $\sigma^+$-polarized laser beam in combination with a $\pi$-polarized infrared laser beam at $795$~nm is used to drive a two-photon transition to a Rydberg state. Inset: relevant atomic levels.}
\end{figure}

We detect the presence of a single atom in the dipole trap by its fluorescence light induced by the optical molasses laser beams. The fluorescence photons are collected using the same high numerical aperture lens and detected on an avalanche photo-diode (APD) in single photon counting mode, as shown in Fig.~\ref{fig:setup}. Due to the small trapping volume the number of atoms in the dipole trap is either zero or one \cite{Schlosser01}.

The temperature of the atom in the optical tweezers is about 70~$\mu$K, which was measured using a release-and-recapture technique \cite{Tuchendler08}.

\subsection{Laser system for Rydberg excitation}\label{section_lasersetup}
We use a two-photon transition to excite a $^{87}$Rb atom from its $5s_{1/2}$ ground state to the desired Rydberg level $58d_{3/2}$. We have chosen this particular state for experiments involving two atoms in neighboring traps~\cite{Gaetan09, Wilk10}, because of the existence of a F\"{o}rster resonance \cite{Walker08, Reinhard08} between the two-atom states $(58d_{3/2}, 58d_{3/2})$ and $(60p_{1/2}, 56f_{5/2})$ which enhances the interaction energy between the atoms. The $5p_{1/2}$ excited state of the D1-line of rubidium is the intermediate level. Therefore, we need a laser at 795~nm and a second laser at 475~nm. As depicted in Fig.~\ref{fig:setup}, the 475~nm laser is $\sigma^+$-polarized and travels along the quantization axis which is defined by a magnetic field of 9~G along the $z$-axis. In the chamber we have about 35~mW of 475~nm light available which is focused to a spot of 20~$\mu$m at the position of the atom. The $\pi$-polarized 795~nm laser has a power of about 7~mW, is applied onto the atoms from the side along the $x$-axis and has a waist of 130~$\mu$m. The fast switching (rise-time $10$~ns) of the blue and red laser beams is done using electro-optic modulators.

We have chosen the intensities of the lasers such, that the excitation towards the Rydberg state is fast enough ($\approx$100~ns) to ensure passive phase coherence between the two lasers for the duration of the excitation pulse (see more details below). The lasers are not actively phase-locked and we control only the frequencies of the lasers with respect to the atomic transitions.

The 795~nm laser is an external cavity diode laser (Toptica DL100) which is locked by a feedback loop acting on the grating piezo-electric transducer. As a reference we use the D1-line of $^{87}$Rb, more precisely the transition between levels  $(5s_{1/2}, F=2)$ and $(5p_{1/2}, F=2)$. The lock scheme relies on a saturated absorption spectroscopy. The error signal is generated using a lock-in technique by modulating the laser diode current at a frequency of 80~kHz. The frequency of the laser is then additionally shifted by a detuning $\Delta$ towards the blue of the transition with an acousto-optical modulator to avoid populating the intermediate state.

The 475~nm laser is a frequency doubling system (Toptica TA-SHG). We stabilize the fundamental wavelength of the 950~nm master laser, as well an external cavity diode laser, by feedback onto its grating piezo-electric transducer. Here, we use a Fabry-Perot cavity to fix the 950~nm laser relative to the 795~nm laser, which gives us more freedom in the choice of the Rydberg state we are addressing. Both, the 795~nm and the 950~nm laser, are coupled to the permanently scanning cavity which has a finesse $F\approx 100$ and a free spectral range of 300~MHz. The cavity is continuously sweeped with a scan frequency of about 100~Hz across the transmission peaks of both lasers.
Our locking scheme relies on the stabilization of the separation between these two transmission peaks in the sweep. We use a home-made circuit to convert the time difference into a voltage. When the signal on the photodiode behind the cavity exceeds for the first time a threshold value, an integrator is started and accumulates a voltage until the photodiode signal exceeds the threshold value for the second time. Subtracting this value with a set voltage generates an error signal, which is treated and fed back onto the 950~nm laser.

This locking technique is only suitable for compensating long-term drifts of the laser, since the bandwidth of the correcting signal is low and depends on the frequency of the Fabry-Perot cavity sweep. Moreover this method relies on the fact that the cavity piezo responds linearly, that the sweep has a very good stability in frequency and that the cavity has slow drifts. Despite those constraints we were able to stabilize the 795-m and 950-nm lasers with respect to each other to about 4~MHz for the duration of $\sim 1$ hour. We also checked the intrinsic short term stability of the 950~nm master laser using a self-heterodyne technique. This consists in observing a beat note of the laser with itself shifted in frequency by 80~MHz and delayed by $\sim 7~\mu$s using a 1.3~km long fiber. We have measured a FWHM of the beat signal of 210~kHz, leading to a linewidth of the 950-nm laser of 105~kHz~\footnote{We assume that the corresponding linewidth of the light generated at 475 nm is twice larger, although we could not measure it directly.  }. Using the same technique, we have measured a linewidth of the 795~nm laser of 600~kHz, which is larger because of the current modulation used to lock it on the atomic transition.

However, this method does not allow to set the wavelength of the 950~nm laser absolutely, since we do not know a priori which longitudinal mode of the cavity we address. We use a wavelength meter with an absolute accuracy of 10~MHz (HighFinesse-\r{A}ngstrom WS-U) to set the frequency of the laser close to the desired frequency. The locking scheme described above has been used for all data shown in this paper and for the data discussed in \cite{Wilk10}. For the data in \cite{Gaetan09} we used an even simpler method: the HighFinesse-\r{A}ngstrom wavemeter provides a PID control option that generates an error signal which can be used to stabilize the laser to any desired frequency with precision of 2~MHz. Automatic recalibration of the wavemeter every 10~s using the 795~nm laser locked to $^{87}$Rb avoids long term drifts of the wavemeter. The limiting factor of this method is the absolute accuracy of the wavemeter, which is estimated around 5~MHz on the time scale of half an hour.

\subsection{Alignment of the 475~nm laser}
\begin{figure}
\includegraphics{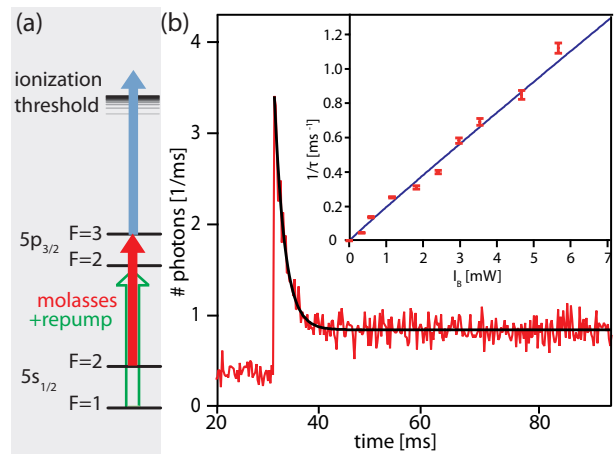}
\caption{\label{fig:ionization} 
(Color online)
Procedure used to align the 475~nm laser onto the single atom. (a) To create population in the $5p_{3/2}$ excited state, the atom is driven by the molasses beams and the repumper. From there, the 475~nm laser photo-ionizes the atom with a rate proportional to its intensity. (b) Fluorescence signal of a single atom averaged over 100 experimental runs. At t=31~ms the atom is exposed to the optical molasses and the 475~nm laser simultaneously. The atom starts fluorescing on the D2-line, is photo-ionized from the 5$p_{3/2}$ state by the 475~nm laser and consequently lost from the trap. The solid line is an exponential fit to the data and gives the photo-ionization rate $1/\tau$. The inset demonstrates the linear dependence of the inverse of the decay time constant as the function of the blue laser intensity. 
}
\end{figure}

In order to reach a high two-photon Rabi frequency with the limited power of the 475~nm laser available, we need to maximize the intensity of the 475~nm light focused to a spot size of about $20~\mu$m at the position of the atom. For this purpose we take advantage of the fact that light at $475$~nm photo-ionizes rubidium atoms from the $5p_{3/2}$-state, see Fig.~\ref{fig:ionization}(a) for excitation scheme, and that the photo-ionization rate is proportional to the light intensity.

After we have prepared a single atom in the tweezers, we turn off the molasses beams and the atom is trapped in the dark. After 31~ms from the start of the sequence we illuminate the atom with both, the 475~nm laser beam and the optical molasses beams. The optical molasses creates population in the excited $5p_{3/2}$-state from where the 475~nm light photo-ionizes the atom which is consequently lost from the trap. Fig.~\ref{fig:ionization}(b) shows a typical trace of the observed fluorescence signal averaged over 100 experimental runs. We observe a peak in the fluorescence when the lasers are switched on which decays exponentially to a value corresponding to the background of the molasses beams and indicating that the atom is lost. A fit by an exponential decay gives the photo-ionization rate $1/\tau$, which depends on the applied powers and on the beam alignments. We verified that the inverse of $\tau$ is proportional to the intensity of the 475~nm laser, see the inset of Fig.~\ref{fig:ionization}(b). Therefore, this method allows us to measure the beam waist of the 475~nm laser  using the single atom as a probe. Keeping the laser power fixed and evaluating the photo-ionization rate for different positions of the laser beam, we find $w_{x}=22(1)~\mu$m and $w_{y}=19(1)~\mu$m which correspond well to the values we expect for our optical setup. The laser beam is then centered onto the atom using the same method.

\subsection{Estimate of the photo-ionization cross section}
The photo-ionization rate can also be used to estimate the photo-ionization cross section $\sigma$ of $^{87}$Rb from the 5$p_{3/2}$ state. The relation between $\sigma$ and $1/\tau$ is given by \cite{Gabbanini97}
\begin{equation}
\label{eq:time_constant}
 \frac{1}{\tau}=f\frac{I_{\rm B}\,\sigma}{h\,\nu_{\rm B}}~,
\end{equation}
where $I_{\rm B}$ and $\nu_{\rm B}$ are the intensity and the frequency of the 475~nm laser, and $f$ is the population in the $5p_{3/2}$ state. The population $f$ depends strongly on the parameters of the optical molasses  and is given by:
\begin{equation}
f=\frac{1}{2}\frac{c^2_1~\Omega^2_{\rm mol}/2}{\delta^2_{\rm mol}+\Gamma^2/4+c^2_2\Omega^2_{\rm mol}/2}\ ,
\end{equation}
where the coefficients $c_1$ and $c_2$ are connected to Clebsch-Gordan coefficients~\cite{Townsend95} and have values of $c^2_1=c^2_2=0.73$~\footnote{Although these values were originally measured for Cs atoms, we use them for $^{87}$Rb, because for the stretched states the Clebsch-Gordan coefficients are the same as for Cs.}. Including the light-shift due to  the dipole trap, the detuning of the optical molasses is $\delta_{\rm mol}/2\pi=100$~MHz.
The total Rabi frequency on the cooling transition of the molasses is given by $\Omega_{\rm mol}= \frac{d}{\hbar}\sqrt{\frac{2I_{\rm mol}}{c \varepsilon_0}}$, where 
$d=3.58\times 10^{-29}$~C$\cdot$m is the transition dipole matrix element \cite{Steck08}, and $I_{\rm mol}$ is the total intensity due to the optical molasses beams at the position of the atom. In our case, the estimation of this intensity can be done only approximately, since it is difficult to determine the position of the single atom within the beams. We estimate the population in the $5p_{3/2}$ state $f$ to be between 1 and 8~\%. This uncertainty contributes the most to the final uncertainty of the ionization cross section.


From the fit in Fig.~\ref{fig:ionization}(b) we extract $\tau=2.03(9)$~ms which was taken for a power of $7.4$~mW of the 475~nm laser. The values of the ionization cross section obtained for the two extreme populations $f$
of the 5$p_{3/2}$ state are
\begin{equation}
0.2\times10^{-17}~\mathrm{cm^2} \leq \sigma \leq 1.6\times10^{-17}~\mathrm{cm^2}\ .
\end{equation}
Despite our uncertainty on $f$  these values agree with theoretically predicted cross sections, which are between $1.25\times10^{-17}~\mathrm{cm^2}$ and $1.4\times10^{-17}~\mathrm{cm^2}$ from $5p$-states to the ionization threshold \cite{Aymar84}, and with  the experimentally measured values from the literature: $1.48(22)\times 10^{-17}~\mathrm{cm^2}$ at 476.5~nm (see~\cite{Gabbanini97} and references therein).

\subsection{Detection of Rydberg atoms}

We detect a successful excitation towards the Rydberg state by a loss of the atom. Our detection method is based on the fact that Rydberg atoms are not trapped in our dipole trap, the light at 810~nm being even slightly anti-trapping for them. For  a Rydberg atom the electron can almost be considered as a free particle and therefore the Thomson model can be applied to calculate the polarizability of the atom~\cite{Saffman05}:
\begin{equation}
\alpha=-\frac{e^2}{m_e \varepsilon_0 \omega^2}< 0\ ,
\end{equation}
where $e$ is the charge of the electron, $m_e$ is the mass of the electron, and $\omega/2\pi$ is the frequency of the applied laser. Since the light shift is proportional to $-\alpha$, and in our case $\alpha < 0$, the light shift is positive for all Rydberg atoms, as long as we are far from any atomic resonance between the Rydberg state and other levels. Since the positive light shift is very small (only $\approx 1$~MHz), the Rydberg atoms leave the trapping region on a time scale that corresponds to their velocities at the temperature of $70~\mu$K. We have estimated the lifetime of a Rydberg atom in the dipole trap by exciting it to the Rydberg state (see section~\ref{section_Rabi}) and measuring the probability to drive it back to the ground state after a given time using a second laser pulse.
After $10~\mu$s the atom has left the trapping region, which is  faster than the radiative decay time back to the ground state which is on the order of $200~\mu$s \cite{Gallagher94}. Moreover, an atom can be photo-ionized by black-body radiation ($85~\mu$s). Photo-ionization by the dipole trap beam ($1.4~$ms~\cite{Saffman05}) is negligible on these timescales.

\section{Rydberg state spectroscopy of a single atom}\label{section_spectroscopy}
In this section we discuss how the desired Rydberg line is initially localized in the spectrum. Then we explain
the optical pumping in state $|5s_{1/2}, F=2, m_F=2\rangle$. From this initial state a two-photon spectroscopy of the $58d_{3/2}$ level is performed on a single atom.

\subsection{Localizing the Rydberg lines}
In order to initially locate the $58d$-line and to have a proof of principle of the detection of Rydberg atoms, we perform a spectroscopy of Rydberg levels using a two-step excitation of the atoms via the $5p_{3/2}$ level. For this particular measurement we use a two-step excitation at 780~nm and the blue laser is tuned to 480~nm. We run the optical tweezers with the optical molasses (780~nm) continuously on and observe the fluorescence from single atoms entering and leaving the trap continuously on the APD while we scan the frequency of the blue laser. When its frequency hits one of the Rydberg lines, the loading of the dipole trap is strongly perturbed, and the atomic fluorescence signal on the APD drops down to the background level.

We have observed $s$, $p$, $d$ and $f$ lines for $n$ ranging from $54$ to $68$, which have a typical separation of $5-30$~GHz. The positions of the lines found with this method coincide within  $\sim100$~MHz with respect to the calculated values, which were calculated using the quantum defect theory~\cite{Gallagher94}. According to the selection rules, only $s$ and $d$ lines should be observable in a two step excitation. The presence of $p$ and $f$ lines indicates the existence of a residual electric field at the position of the atom.

With this spectroscopy method, we identified the $58d_{3/2}$ line, which we use in further experiments.

\subsection{Initial state preparation: optical pumping}
In all further experiments we perform a direct two-photon excitation of a single atom trapped in the tweezers from its ground state $|5s_{1/2}, F=2, m_F=2\rangle$. The atom is optically pumped into this particular state during $600~\mu$s using two laser beams: $\sigma^+$-polarized light resonant with the atomic transition $|5s_{1/2}, F=2\rangle$ to $|5p_{3/2}, F=2\rangle$ and repumping light resonant with the transition $|5s_{1/2}, F=1\rangle$ to $|5p_{3/2}, F=2\rangle$. To ensure that there is no population left in the $F=1$ ground state, the repumping light is switched off $1~\mu$s later than the pumping light. The efficiency of the optical Zeeman pumping to the state $|5s_{1/2}, F=2, m_F=2\rangle$ is about $95\%$, which was checked by measuring the visibility of Rabi oscillations driven by Raman transitions between the $|5s_{1/2}, F=1, m_F=1\rangle$ and $|5s_{1/2}, F=2, m_F=2\rangle$ hyperfine ground states~\cite{Jones07}.

\subsection{Spectroscopy of the $58d_{3/2}$ level}

\begin{figure}
\includegraphics{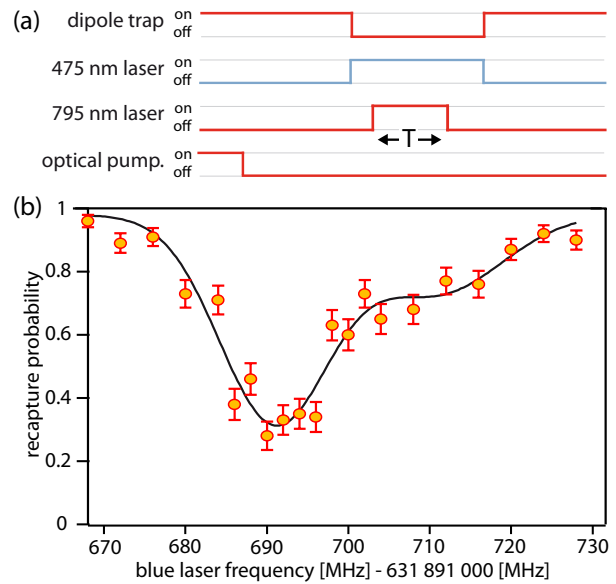}
\caption{\label{fig:spectrum} 
(Color online) Two-photon spectroscopy of the $58d_{3/2}$ line. (a) The blue laser is switched on for $600~$ns. The precise timing of the two-photon transition is controlled by the duration $T$ of the infrared square pulse with a rise time less than 10~ns. The frequency scan is done by varying the frequency of the blue light. (b) The corresponding spectrum taken for $T=60~$ns and $(\Omega_{\rm R},\Omega_{\rm B},\Delta)/2\pi =(255,24,400)$~MHz, see Section~\ref{section_Rabi}. Each point corresponds to 100 repetitions of the experiment. The solid line is a double-Gaussian fit to the experimental points.}
\end{figure}

Figure~\ref{fig:spectrum}(a) shows the time sequence used for this experiment. About $1~\mu s$ after the optical pumping, we switch on the 475~nm laser beam. We also switch off the dipole trap for 600~ns to avoid light shifts due to the dipole trap laser light during the Rydberg excitation. During this time window the 795-laser is switched on for $T=60$~ns to coherently excite the free atom to the Rydberg state, as will be detailed in Section~\ref{section_Rabi}. After this excitation the dipole trap is switched on again, which results in the loss of the atom if it was in the Rydberg state and in its recapture in the optical tweezers otherwise. The frequency scan is performed by changing the absolute frequency of the 475~nm laser. For every point we repeat the experiment 100 times.

Figure~\ref{fig:spectrum}(b) presents the resulting two-photon spectrum. We observe a double peak structure. The two peaks have a separation of $20(1)$~MHz from each other. We attribute the more pronounced dip to the transition to the $|58d_{3/2}, F=3, m_F=3\rangle$ state. The width (FWHM) of this main peak is $16(1)$~MHz. It is compatible with the expected width ($13.3$~MHz) using a square pulse with duration $T=60$~ns convoluted with the contribution from experimental imperfections ($6$~MHz, see section~\ref{section_Rabi}). These are mainly fluctuations of the excitation laser frequencies and intensities as well as the rise- and fall-time of the square pulse. The center of the line is at  $\nu_{\rm B}^{\rm exp}=631\,891\,691(10)$~MHz, where the error bar results from the absolute precision of the wavemeter. We calculate the position of this line to be at $\nu_{\rm B}^{\rm calc}=631\,891\,657(6)$~MHz using quantum defect theory and taking into account the light-shifts induced by the 795~nm and the 475~nm  lasers and the Zeeman shifts. The difference between the calculated and measured frequencies of $34(12)$~MHz can have several reasons: an uncertainty of the light-shift induced by the 795~nm laser beam itself during the excitation ($\approx 10$~MHz), or the absolute precision of the wavemeter ($\approx10$~MHz). Moreover, stray electric fields, which we can not control on our setup, would also induce a shift. However, the calculated Stark shift ($-240\ {\rm MHz}/({\rm V/cm})^2)$) does not have the correct sign to explain the observed frequency difference.

The origin of the satellite dip is not entirely clear at this stage and further investigation is required. A possible candidate is the imperfect optical pumping in the Zeeman state that leaves population in $|5s_{1/2}, F=2, m_F=1\rangle$. From that state the $|58d_{3/2}, F=2, m_F=2\rangle$ and $|58d_{3/2}, F=3, m_F=2\rangle$ levels can be excited by the two-photon transition. Since the $|5s_{1/2}, F=2, m_F=1\rangle$ state is light-shifted by the 795~nm laser by a smaller amount than the $|5s_{1/2}, F=2, m_F=2\rangle$ state, due to a smaller Clebsch-Gordan coefficient connecting it to the corresponding Zeeman state in level ($5p_{1/2}$), we expect the side dip to be at a higher frequency of the 475~nm laser, as we observe it on the measured spectrum. This is also confirmed by the fact that changing the detuning $\Delta$ from the intermediate state from positive to negative detuning changes the position of the side dip from the blue to the red side of the Rydberg line. 
However, the area of this side dip seems to be too large to be explained only by about 5\% imperfect optical pumping. Another possible effect is the non-perfect polarization of the excitation laser at 475~nm.


\section{Rabi oscillation between ground and Rydberg state}\label{section_Rabi}

Once the  state $|58 d_{3/2}, F=3, m_F=3\rangle$  is localized in the spectrum, we set the frequency of the blue laser to the center of the line.  By varying the pulse duration $T$ of the 795~nm laser we observe Rabi oscillations  between the ground state $|5s_{1/2}, F=2, m_F=2\rangle$ and the Rydberg state $|58 d_{3/2}, F=3, m_F=3\rangle$, demonstrating the coherence of the two-photon excitation. We discuss the experimental results of measurements with different Rabi frequencies of the red (795~nm) and the blue (475~nm) laser, and for different detunings from the intermediate state $\Delta$. The measurements are compared to simulations that we performed which include intensity and frequency fluctuations present in our system.

\subsection{Experimental results}

\begin{figure}
\includegraphics{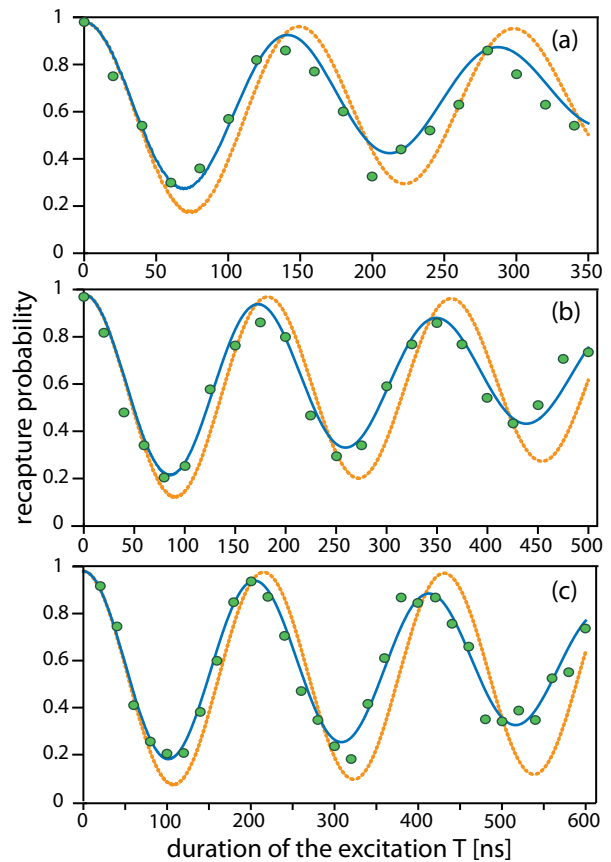}
\caption{\label{fig:Rabi} 
(Color online)
Coherent Rydberg excitation of a single atom to the level $|58d_{3/2},F=3,m_F=3\rangle$ for different experimental parameters:
(a) $(\Omega_{\rm R},\Omega_{\rm B},\Delta)/2\pi =(255,24,400)$~MHz,
(b) $(\Omega_{\rm R},\Omega_{\rm B},\Delta)/2\pi =(250,28,600)$~MHz, and
(c) $(\Omega_{\rm R},\Omega_{\rm B},\Delta)/2\pi=(80,70,600)$~MHz.
Each point corresponds to 100 repetitions of the experiment. The blue line is the result of a Monte-Carlo simulation of the dynamics of a five-level system, which includes a decay from the intermediate state, fluctuations of the power and of the frequency of the lasers and imperfection of the optical pumping. The dotted orange line shows for comparison the results of simulations with the same parameters of $\Omega_{\rm R}$, $\Omega_{\rm B}$, and $\Delta$ but without fluctuations.}
\end{figure}

We have recorded the Rabi oscillations between ground and Rydberg states for different values of the red and blue Rabi frequencies $\Omega_{\rm R}$, $\Omega_{\rm B}$, and the detuning $\Delta$. In Fig.~\ref{fig:Rabi} the recapture probability of the single atom is plotted versus the duration $T$
of the excitation pulse. Each point corresponds to 100 repetitions of the experiment. Fig.~\ref{fig:Rabi}(a) shows the measurement for $(\Omega_{\rm R},\Omega_{\rm B},\Delta)/2\pi=(255,24,400)$~MHz. For these parameters we reach a maximal excitation probability towards the Rydberg state of $\approx 70~\%$, and the oscillation is strongly damped. A fit to the data by the function $A-B\,{\rm e}^{-\frac{T}{\tau}}\cos\Omega T$ (not shown) yields  a two-photon Rabi frequency of $\Omega/2\pi=7.0(1)$~MHz and $\tau = 480$~ns. The solid and the dotted line are results of simulations that will be discussed in the next subsection.

We attribute the damping mainly to spontaneous emission from the intermediate $5p_{1/2}$ state, which destroys the coherence. A second measurement, shown in Fig.~\ref{fig:Rabi}(b), with about the same Rabi frequencies of the lasers $(\Omega_{\rm R},\Omega_{\rm B})/2\pi=(250,28)$~MHz but with a larger detuning $\Delta/2\pi=600$~MHz already improved the excitation probability up to $80\%$. The fit yields $\Omega/2\pi=5.8(1)$~MHz and a longer damping time $\tau = 560$~ns. The damping rate could even be lowered further by reducing the Rabi frequency of the red laser significantly to $\Omega_{\rm R}/2\pi=80$~MHz and by increasing the one of the blue laser to $\Omega_{\rm B}/2\pi=70$~MHz (stronger focusing of the blue laser and improvement of the optical setup). The fit gives  $\Omega/2\pi=4.9(1)$~MHz and $\tau = 912$~ns. But even at these parameters the excitation probability to the Rydberg state does not exceed 85~\%. To understand better the underlying mechanisms for the damping and the limited excitation probability, we have modeled the system including experimental imperfections as will be discussed in the following subsection.

\subsection{Comparison to numerical simulations}

In our numerical calculation we include five atomic states, which are labeled in the level scheme of Fig.~\ref{fig:setup}: in the $5s_{1/2}$ ground state we define $|\!\downarrow\rangle = |F=1, m_F=1\rangle$,  $|\!\uparrow\rangle = |F=2, m_F=2\rangle$ and  $|g\rangle = |F=2, m_F=1\rangle$. We call the intermediate state  $|p\rangle = |5p_{1/2}, F=2,  m_F=2\rangle$  and the Rydberg level is  $|r\rangle$. We take into account the spontaneous emission from the intermediate level $|p\rangle$ to the ground state with a rate of $\Gamma/2\pi =5.75$~MHz and the corresponding branching ratios of $1/2$ to state $|\!\downarrow\rangle$, $1/3$ to $|\!\uparrow\rangle$, and $1/6$ to $|g\rangle$~\cite{Steck08}, as well as the decay from the Rydberg state to the intermediate level with rate $\gamma/2\pi=4.8$~kHz.

The time evolution of the density matrix $\rho$ is governed by the master equation
\begin{equation}
\dot{\rho}=-\frac{i}{\hbar}\left[\hat{H},\rho \right]+\mathcal{L}\ ,
\end{equation}
where the Hamiltonian $\hat{H}$ and the Liouville $\mathcal{L}$ operators are given in the basis $\{|\!\downarrow\rangle,|g\rangle,|\!\uparrow\rangle,|p\rangle,|r\rangle\}$ by~\cite{Fleischauer05}
\[
\hat{H}=\hbar \left( \begin{array}{ccccc} -\omega_{\rm HF} & 0 & 0 & 0 & 0 \\ 0 & 0 & 0 & 0 & 0 \\ 0 & 0 & 0 &
\Omega_{\rm R}/2 & 0 \\ 0 & 0 & \Omega_{\rm R}/2 & -\Delta & \Omega_{\rm B}/2 \\ 0 & 0 & 0 & \Omega_{\rm B}/2 & -\delta \\
\end{array} \right),\]
and
\[\mathcal{L}= \left( \begin{array}{ccccc}
\frac{\Gamma}{2}\rho_{pp} & 0 & 0 & -\frac{\Gamma}{2}\rho_{\downarrow p} & -\frac{\gamma}{2}\rho_{\downarrow r} \\
0 & \frac{\Gamma}{6}\rho_{pp} & 0 & -\frac{\Gamma}{2}\rho_{gp} & -\frac{\gamma}{2}\rho_{gr} \\
0 & 0 & \frac{\Gamma}{3}\rho_{pp} & -\frac{\Gamma}{2}\rho_{\uparrow p} & -\frac{\gamma}{2}\rho_{\uparrow r} \\
-\frac{\Gamma}{2}\rho_{p \downarrow } & -\frac{\Gamma}{2}\rho_{pg} & -\frac{\Gamma}{2}\rho_{p\uparrow } & \gamma\rho_{rr}-\Gamma \rho_{pp} &-\frac{\Gamma +\gamma}{2}\rho_{pr}  \\
-\frac{\gamma}{2}\rho_{r\downarrow } & -\frac{\gamma}{2}\rho_{ rg} & -\frac{\gamma}{2}\rho_{r\uparrow } & -\frac{\Gamma+\gamma}{2}\rho_{rp}  &  -\gamma \rho_{rr} \\
\end{array} \right).\]
Here $\omega_{\rm HF}/2\pi = 6.834$~GHz is the hyperfine ground state splitting  frequency. We include the power and frequency fluctuations of the two lasers by  applying a Monte Carlo simulation. We assume that the power fluctuations of the red and the blue laser are both Gaussian distributed, and are  2.5\% and 5\% (FWHM), respectively, as measured independently. We note that these intensity fluctuations also lead to a fluctuation of the frequency of the transition  due the light-shift induced by the lasers on this two-photon transition, respectively $\Omega^2_{\rm R}/4 \Delta$ on the ground state  and $\Omega^2_{\rm B}/4 \Delta$ on the Rydberg state. To take into account the frequency fluctuations of the lasers we assume a Gaussian distribution of the two-photon detuning with respect  to the transition $\delta$. This procedure would also include a potential variation in the Rydberg transition frequency by the Stark effect due to a fluctuating  stray electric field. We take into account the efficiency of the optical pumping of 95~$\%$ by assuming that the remaining 5~\% are in state $|g\rangle$. Finally, we multiply the results of the simulation by 0.98 to account for the measured probability to recapture the atom at the end of the sequence in the absence of laser excitation. We average over 100 evolutions of the master equation, and the solution  is shown for different values of $\Omega_{\rm R}$, $\Omega_{\rm B}$, and $\Delta$  in Fig.~\ref{fig:Rabi} as blue solid line. In order to have an idea of how the fluctuations influence the result we have added on the graphs of Fig.~\ref{fig:Rabi} the dotted orange line which corresponds to the solution of  the model, without including the frequency and power fluctuations of the lasers.

The simulations are in good agreement with the  measured data. For the simulations  shown in blue in Fig.~\ref{fig:Rabi} we assumed fluctuations (FWHM) of the  two-photon detuning $\delta$ of 6~MHz in (a), 4.5~MHz in (b) and 4~MHz in (c). Comparing them to the results without fluctuations, we identify two main factors limiting the population transfer to the Rydberg state: the spontaneous emission from state $|p\rangle$ and the fluctuations of laser frequency and intensity. The influence of the spontaneous emission can be reduced by decreasing the ratio $\Omega^2_{\rm R}/\Delta^2$, resulting in an increase of the contrast of the Rabi oscillations, as shown in Fig.~\ref{fig:Rabi}(b) and (c). Moreover, we estimate from our simulations, that if we would reduce the laser frequency fluctuations to $1~$MHz, we could increase the Rydberg excitation efficiency to $93~\%$. 
We finally note that the frequency fluctuation used in the model are compatible with  the one  estimated from measurements on the laser system (see  section~\ref{section_lasersetup}). At the present stage of the experiment we therefore have no evidence of a fluctuation of the transition frequency due to  stray electric fields.

As a last comment, we conclude from the good agreement between the data and the model, that the result of the simulation is compatible with $100\%$ efficiency of the Rydberg state detection we use on the experiment.

\section{Conclusion}
We perform coherent excitation of single atoms to the Rydberg state $|58d_{3/2},F=3,m_{F}=3\rangle$ using a two-photon transition and observe  Rabi oscillations with a high contrast. The observed Rabi frequency is about 7~MHz, which relaxes partially the requirements to the frequency stability of the laser system. The largest population transfer to the Rydberg state observed is $80~\%$. The good agreement between the model and the data leads us to conclude that this maximal transfer efficiency  is limited by the spontaneous emission from the intermediate state and the frequency stability of the excitation lasers. In future experiments the population of the intermediate level can be reduced by increasing the detuning from this level and using higher power in the blue beam to keep the two-photon Rabi frequency high.

\begin{acknowledgments}
We would like to thank Matthieu Viteau and Amodsen Chotia for their contributions at an early stage of the experiment. We acknowledge support from IARPA, the European Union through the Integrated Project AQUTE and the ERC starting grant ARENA, and the Institut Francilien des Atomes Froids (IFRAF). A. Ga\"{e}tan and C. Evellin are supported by a DGA fellowship. Y. Miroshnychenko and T. Wilk  are supported by the IFRAF.
\end{acknowledgments}

\end{document}